\documentclass{ws-procs9x6}

\newcommand{\beqn}{\begin{eqnarray}}
\newcommand{\eeqn}{\end{eqnarray}}
\newcommand{\be}{\begin{equation}}
\newcommand{\ee}{\end{equation}}

\def\st{Stueckelberg~}
\def\s1{$s_{\alpha}$}
\def\s2{$s_{\gamma}$}
\def\s3{$s_{\delta}$}
\def\c1{$c_{\alpha}$}
\def\c2{$c_{\gamma}$}
\def\c3{$c_{\delta}$}
\def\y{Y_{\phi}}

\def\beq{\begin{equation}}
\def\eeq{\end{equation}}
\def\beqn{\begin{eqnarray}}
\def\eeqn{\end{eqnarray}}
\def\st{Stueckelberg\ }

\def\s1{$s_{\alpha}$}
\def\s2{$s_{\gamma}$}
\def\s3{$s_{\delta}$}
\def\c1{$c_{\alpha}$}
\def\c2{$c_{\gamma}$}
\def\c3{$c_{\delta}$}
\def\s{Stueckelberg~}

\newcommand{\mathsym}[1]{{}}

\begin{document}

\title{Dark Matter  in SUGRA, Strings and Branes}

\author{Pran Nath }

\address{Department of Physics, Northeastern
University, Boston, MA 02115, USA\\
$^*$E-mail:  nath@neu.edu}

\begin{abstract}
A brief review is given of dark matter in SUGRA, strings and branes.
For SUGRA models the implications of Yukawa coupling unification 
 on dark matter are discussed in the light of $g-2$ and $b\to s\gamma$ 
constraints. A brief discussion is given of the dark matter in orbifold
string compactifications under constraints of modular invariance and
radiative breaking of the electroweak symmetry.   Finally a new candidate for dark matter -
an extra-weakly interacting massive particle or an XWIMP- is discussed.
Such dark matter can arise in a wide class of models,
 including the Stueckelberg extensions of MSSM,  in U(1) extensions
 of MSSM with off diagonal kinetic energy, and possibly in a wider class
 of models which may have a string/D-brane  origin.  Satisfaction of the 
 relic density of XWIMPs consistent with WMAP 
 is also discussed.
\end{abstract}
\bodymatter
\section{Introduction}
 In this talk we  give a brief overview 
 of the leading candidate for cold dark matter \cite{Clowe:2006eq} 
 in a  broad class of models
 which includes the supergravity unified models (SUGRA),   string  models and
 brane models. In mSUGRA\cite{msugra}  the neutralino\cite{Goldberg:1983nd}
arises  as  the least massive supersymmetric particle (LSP) over  a broad  region
 of the parameter  space\cite{Arnowitt:1992aq}, 
 and with R parity it can be a candidate  for 
 cold dark matter.
 The nature  of dark matter  depends critcally on the type
 of soft breaking and it is  this aspect that differentiates  the nature of dark
 matter in SUGRA models, vs dark matter  in strings  and branes.  
 For instance, in orbifold compactifications  of the heteoric string, 
 the constraints of modular invariance play an important role in the 
 nature of soft breaking, and hence on the nature of neutralino dark matter.
 Aside from the neutralino, even in supersymmetry  there exist  other 
 possible  candidates  such as  the gravitino, and the  sneutrino. Specifially,
 the gravitino possibility has  resurfaced  recently\cite{gravitino}.
In addition other dark matter candidates abound such as the  Kaluza-Klein
 states\cite{kk}  in extra dimensional models, 
 Q balls\cite{Kusenko:2001vu}, as well as a variety of other
  possibilities\cite{Covi:2004rb,Burgess:2000yq,Frampton:2006va}.
 To this list we will add  a new candidate - an extra
 weakly interacting massive  particle or an XWIMP\cite{Feldman:2006wd}. 
 Such a  particle  can arise in a  wide class of models including the
 \st extensions of MSSM\cite{kn1,kn2,kn3,Feldman:2006ce,Feldman:2006wb},  
  the $U(1)_X$ extensions with off diagonal kinetic
 energy terms\cite{Holdom:1985ag,Kumar:2006gm,Chang:2006fp,Feldman:2006wd},
  and possibly  in a broader class of models with string/D-brane origins. 
 
 The outline of the rest of the paper is as follows. In Sec.(2) we 
 give an overview of dark matter in SUGRA models with 
  focus on inclusion of Yukawa coupling 
 unification. It  is known that  in this case  the constraints of $b\to s+\gamma$,
 and the sign of the $\mu$  parameter  play a central role  in the analysis.
 Since the sign of $\mu$ is closely tied  to  the sign of the  supersymmetric 
 contribution to  $g_{\mu}-2$,  the analysis  in this  case is highly constrained.
 In Sec.(3) we  discuss dark matter in heterotic  string models,
 and point out 
  that here $\tan\beta$ in no longer a free  parameter but is
 determined by the twin constraints  of radiative breaking of the  
 electroweak symmetry and by the constraints  of  modular  invariance.  
 In Sec.(4) we  discuss the new candidate for  dark  matter- an 
 extra-weakly interacting  massive particle, and show that  despite  its  extra
 weak interactions, it  is possible to satisfy the WMAP relic density
 constraints.

\section{Dark matter in SUGRA unifiication} 
Extensive  investigations of the relic density  in SUGRA models exist  in the literature
(for recent works see\cite{recentdark,d1,d2})  and  
for the mSUGRA case this implies exploration of the parameter space spanned by 
the four conventional parameters: $m_0, m_{1/2}, A_0, \tan\beta$.  Significant regions
exist where WMAP constraints  can be  satisfied and these  regions can be  broadly
labeled as the stau co-annilation region, where coannhiliation between the LSP and
the stau produces  relic densities  consistent with WMAP,  the resonance  region 
where relic density constraint 
 is satisfied due to  the Higgs poles, and  the hyperbolic branch (HB)\cite{hb} 
where the relic density  is satisfied  due to a relatively large  higgsino component of
the  LSP.  These analyses are  very sensitive  to the nature  of  soft  breaking
and thus  the inclusion of non-universalires  in the  soft breaking produce significant
effects in the analysis.  Non-universalites  can appear  in a variety of ways  but these
must be consistent with the flavor changing neutral current constraints.  
Such constraints can be respected  by inclusion of non-universalites  in the Higgs sector
and in the gaugino sector, and  several  analyses   exist  where the  Higgs sector\cite{Nath:1997qm}
and the  gaugino 
sector\cite{Corsetti:2000yq,Birkedal-Hansen:2001is,Belanger:2004ag,Cerdeno:2004zj,Chattopadhyay:2003yk,Bottino:2004qi}
non-universaliites have been included in dark matter analyses.  In addition to
the above  analyses of dark matter are  also sensitive  to CP phases and we discuss this later. 

\begin{figure}[h]
\hspace*{-.2in} \centering
\includegraphics[width=8cm,height=6cm]{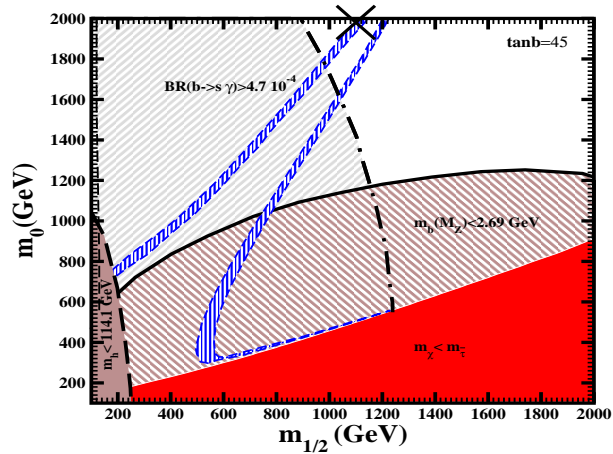}
\caption{Analysis of the neutralino
relic density with the $b-\tau$ Yukawa unification constraints 
in the $m_0-M_{1/2}$ plane when the soft terms are universal and real 
with $\mu<0$ ($\theta_\mu=\pi$), $A_0=0$, $m_t=178$~GeV,and 
$tan\beta=45$.  
Areas contoured by the dashed line has a neutralino relic density which is
inside  the WMAP bounds\cite{Bennett:2003bz,Spergel:2003cb}.
The area above the solid line predicts $m_b(M_Z)>3.10$ GeV,
while the area inside the dashed (dot-dashed) line is excluded by the lower bound
on $m_h$ (the upper bound on $BR(b\to s+\gamma)$. On the lower  dark  area  
$m_{\chi}>m_{\tilde \tau}$  while  on the upper side  EWSB is not achieved. The thin
dashed line  indicates $m_{\chi^+}=103$ GeV. 
Taken from Ref. \cite{gins}} \label{fsetx}
\end{figure}

For the remainder of this section we will focus on the analysis of dark matter
including the effects of
Yukawa coupling  unification. Thus in  many unified  models the b and $\tau$ Yukawa couplings are related at high scales, e.g., 
$h_b\simeq h_{\tau}$. These are evolved down to the electroweak scale  and constrained by
experiment
$m_{\tau}=1.7463 ~{\rm GeV}$,  and 
$2.69 {\rm GeV} < m_b(M_Z) <3.10 {\rm GeV}$. 
In some models one extends the above to a full Yukawa unification 
$h_{\tau} \simeq h_b\simeq h_t$.  
We note in passing that while $b-t-\tau$ unification in $SO(10)$ models with 10 plet 
of Higgs for breaking of the electroweak symmetry requires a large $\tan\beta$, 
a large $\tan\beta$ is not a necessity when the symmetry breaking is achieved 
via alternative schemes (see, e.g.,  Ref\cite{bgns}).

The sign of $\mu$ plays a central role 
in  $b-\tau$ and   $b-t-\tau$ unification.  It is known that the supersymmetric
  contribution to $a_{\mu} =(g_{\mu}-2)/2$ is directly correlated to the sign of
 $\mu$\cite{kosower} and further that one can infer this sign experimentally\cite{Chattopadhyay:2001vx}. 
 A positive $\mu$ is  favored by the  $b\to s\gamma$
  since the parameter space of 
 msugra and of other  models is less stringently constrained by
 it \cite{Nath:1994ci}.  On the other hand $b-\tau$ unfication seems to favor
 a negative $\mu$   \cite{Pierce:1996zz,deBoer:2001xp}.
  This is so because $b-\tau$ unification 
 requires a negative loop contribution to the b quark mass, and  the major contribution
 to this loop comes from the gluino exchange and its sign depends on
  $\mu \tilde m_g$.   Many analyses exist which  have worked to 
 resolve this  problem\cite{Baer:2001yy,bdr,ky,ccnbtau}. 
 One such possibility 
 is to use non-universalities\cite{ccnbtau}. For example,
in $SU(5)$ the gaugino masses transform like the symmetric product of $(24\times 24)$ which
can be expanded as  $1+24+75+200$. 
Now  for the singlet case one gets universality of 
gaugino masses at the GUT scale.  However, if one considers the 24 plet case, then
$M_3, M_2, M_1$ are in the ratio (-2, 3, 1), and one finds a relative  minus sign for the gluino mass term.
This gaugino mass pattern 
 switches the  sign of  $(\mu.m_{\tilde g})$ from positive to negative,  which
  allows one to achieve  a $b-\tau$ unification with a  positive  $\mu$. 
Experimentally, 
the most recent analyses appear  to favor a positive  $\mu$\cite{Stockinger:2006zn}. 
Still we discuss both $\mu$ signs for 
Yukawas unification and dark matter\footnote{For an analysis of dark matter with quasi -Yukawa unifcation see Ref.\cite{Pallis:2003jc}}. 
For positive $\mu$ the analysis of dark  matter  
is given in Refs.\cite{ccnbtau}, while for negative $\mu$ it is given in
Ref.\cite{gins},  and  an exhibition of one case  is given in Fig.(1) 
 using the WMAP relic density  of
Ref.\cite{Bennett:2003bz,Spergel:2003cb}. 
The analysis of Fig.(1) shows that Yukawa unification constraint allows for a satisfaction of the relic
density constraint consistent with  WMAP\cite{Bennett:2003bz,Spergel:2003cb}.
CP phases also have  a strong effect on dark matter\cite{gins,Chattopadhyay:1998wb} 
but here one needs to 
pay attention to the satisfaction of the edm constraints which, however, can be satisfied
even for large phases via the cancellation mechanism\cite{cancellation}. 
In passing we draw attention to the recent  improved analyses  of $b\rightarrow s + \gamma$
which,  as discussed above,  has an  important  effect on  dark matter. These  improved analyses include
the next  to leading order (NLO) corrections enhanced by large $\tan\beta$\cite{bsg,Gomez:2006uv}.    
The most recent analysis of Ref.\cite{Gomez:2006uv} additionally  includes the full arrary of CP violating 
effects and these results will be useful in future dark matter analyses. 
In unified models there is also a strong link between proton stability and 
dark matter\cite{Arnowitt:1998uz, Nath:2006ut} a topic which is beyond the scope of this talk. 
  \begin{figure}[h]
\hspace{-.2in} \centering
\includegraphics[width=8cm,height=6cm]{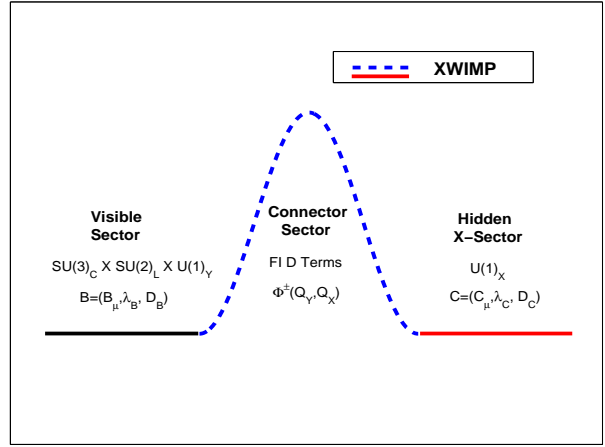}
\caption{An exhibition of the mechanism by which an  XWIMP is generated. 
A neutral fermion from the connector sector merges with a gaugino of the
hidden sector to form an  XWIMP whose  interactions with the
MSSM particles are suppressed.  Taken from Ref.\cite{Feldman:2006wd}}
 \label{connector}
\end{figure}
\section{Dark matter in heretotic string models}
As in SUGRA models,  dark matter in hererotic string models is largely governed by the soft breaking. 
In orbifold  compactifications one typically has a large radius- small radius symmetry, so that
$R \to \alpha'/R$, and more generally an  $SL(2,Z)$ modular invariance  symmetry.  There  are 
many analyses which have looked at soft breaking with modular invariance\cite{brignole,gaillard}
 and their  implications (For a sample
see, Refs.\cite{kane,nathtaylor,Chattopadhyay:2004ct,Binetruy:2003yf} and references therein).  In the analysis of 
Ref.\cite{nathtaylor,Chattopadhyay:2004ct}, the  
 further constraint of radiative breaking of the
electroweak symmetry was utilized. With the twin constraints of  modular invariance and 
radiative breaking, $\tan\beta$ is no longer a free parameter but is determined in terms of 
$\alpha_{string}$  and the remaining  soft parameters.
 Using this constraint an analysis of dark matter   
 for $\mu>0$ with WMAP constraint  implies an  upper limit on sparticle 
masses which lie within reach of the LHC, and further  the neutralino-proton cross sections
  lie within reach of the dark matter detectors\cite{Chattopadhyay:2004ct}.\\
\section{Extra weakly interacting  dark matter}
Recently a new 
candidate for dark matter has been proposed whose couplings with matter are weaker than
weak or extra weak\cite{Feldman:2006wd}.   The mechanism for its generation is exhibited  in Fig.(2), and it depends on three sectors:
a visible  sector  where the particles  of MSSM reside,  a  hidden sector   where
fields do not have any  direct  interactions with the fields in the visible sector
and a third sector\cite{Nath:1996qs}
 which connects  both to the fields of the visible sector and of the hidden sector.
We label this third  sector, the connector sector.   A spontaneous breaking in the connector sector 
produces mixings between the neutralino states in the visible sector and the neutralino states 
in  the hidden sector.
If the LSP of the hidden sector (XLSP) lies  lower than the LSP of the visible  sector,  then the XLSP 
becomes the LSP of the entire  system. This is the XWIMP.
As an example we  consider the $U(1)_X$ Stueckelberg extension
\cite{kn1,kn2,kn3,Feldman:2006ce,Feldman:2006wb}
 where one has the 
$U(1)_X$ gauge fields $C_{\mu}, \lambda_C,  D_C$.
The connector sector is chosen to be a pair
cf chiral fields $\phi^{\pm}$\cite{Dvali:1996rj,kn4} 
which are charged under both  $U(1)_X$ and  $U(1)_Y$.
We add to the mix a Fayet-Illiopoulos term\cite{Fayet:1974jb} 
\beqn
{\cal L_{FI}} = \xi_X D_C+\xi_Y D_B
\eeqn 
Elimination of the D terms then leads to the potential 
\beqn V= \frac{g_X^2}{2} \Big(Q_X|\phi^+|^2 -Q_X|\phi^-|^2
+\xi_X\Big)^2 +\frac{g_Y^2}{2} \Big(\y |\phi^+|^2 -\y |\phi^-|^2
+\xi_Y\Big)^2\  . \eeqn
which on minimization gives    $\langle\phi^+\rangle=0\ , \quad\langle\phi^-\rangle\neq 0$.
 After  spontaneous breaking two new mass parameters emerge. 
 $M_1=\sqrt 2 g_XQ_X <\phi^->, ~~M_2=\sqrt 2 g_YY_{\phi} <\phi^->$, or 
 alternately one can choose the new  parameters to be $M_1$ and
 $\epsilon =M_2/M_1$. 
The hidden sector and the connector sector provide  two additional 
neutralino fields, $\chi_S, \lambda_X$ which together with the 
four neutral fermionic states in the MSSM,
$\lambda_Y,\lambda_3, \tilde h_1, \tilde h_2$ give a set of 
six  Majorana spinors.  In the basis 
 $((\chi_S,\lambda_X);(\lambda_Y, \lambda_3, \tilde h_1, \tilde h_2))$
one finds a $6\times 6$ Majorana mass matrix, whose 
 eigen states we  label 
$ ((\xi_1^0, \xi_2^0) ;( \chi_1^0,\chi_2^0,\chi_3^0,\chi_4^0)) $, 
where $\chi_a^0$  (a=1,2,3,4) are essentially
  the four neutralinos
that appear in MSSM, and  $\xi_{\alpha}^0$, $(\alpha =1,2)$ 
are the new  states.

Following the same procedure used to constrain extra dimensions\cite{Nath:1999fs}
one can put also a constraint on $\epsilon$,   and one  finds  
$\epsilon<0.06$\cite{Feldman:2006wd}. 
The LEP and the Tevatron data put further constraints on the model.\cite{Feldman:2006wd}
reducing further the size of $\epsilon$. 
Because of the smallness of $\epsilon$,  the interactions
of $\xi_{\alpha}^0$ with the visible  sector fields will be suppressed
by an additional factor of $\epsilon$ and thus the interactions of the XWIMP 
with the visible sector will be extra weak. 
Further, if the mass of either $\xi_1^0$ or $\xi_2^0$ is smaller 
than the mass of all of the MSSM particles,  then the XLSP will be the 
LSP of the entire system and hence with R parity conservation, a
candidate for dark matter.  
\begin{figure}[h]
\hspace*{-.2in} \centering
\includegraphics[width=8cm,height=6cm]{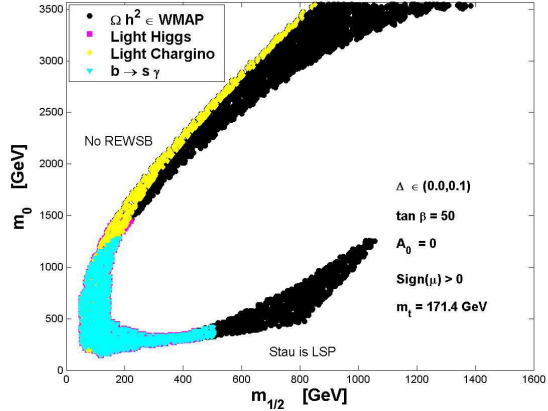}
\caption{The allowed parameter space in the  $m_0-m_{1/2}$ plane,
 under  the $1\sigma$ WMAP3 constraint\cite{Spergel:2006hy}  in extended mSUGRA
 for the case $A_0=0$, $\tan \beta = (30,50)$ (upper,lower),  sign($\mu>0$),
  $m_t=171.4$ GeV,  $m_{1/2}\in (0,1.5)$TeV and $m_0 \in (0,3.5) $TeV, and
$\Delta$ in the range (0.0, 0.1).
  Regions eliminated  by the light chargino mass constraint, by the light
Higgs mass constraint, and by the $b\to s+\gamma$ constraint are
also exhibited. Taken from Ref. \cite{Feldman:2006wd}}
\end{figure}
It was shown in \cite{Feldman:2006wd}
 that a similar situation arises for the  $U(1)_X$ extension of  MSSM with gauge
kinetic energy mixing\cite{Holdom:1985ag,Kumar:2006gm,Chang:2006fp}
involving a  mixing  of the $U(1)_X$  and the $U(1)_B$
gauge field strengths.  Supersymmetric version of this model leads to a form of the neutralino
mass matrix which, although different in form, also produces an XWIMP. Further, one may 
conceive of other models where considerations of the type outlines above lead to an
 XWIMP and  a candidate for cold dark matter. Models of the type discussed here may have 
 a string/D-brane origin\cite{stringorigin}.

 A priori it would appear that an XWIMP would not satisfy the relic density constraints as its  extra
 weak interactions would not  allow for an efficient annihilation of excess CDM in the early universe. 
 However, if  any of the  MSSM particles lie  close to the  XWIMP mass, then the XWIMPS can 
 co-annihilate quite  efficiently and the relic density constraints  can be satisfied. 
   Thus, for example, if the NLSP turns  out to be  
 $\chi^0=\chi_1^0$, and if  $\Delta =(m_{\chi^0}-m_{\xi^0})/m_{\chi^0} >0$, then
 the XWIMP $\xi^0=\xi_1^0$ relic density is given by\cite{Feldman:2006wd} 
 \beqn
 \sigma_{eff}(XWIMP)\simeq \sigma_{eff}^{\chi^0\chi^0}(\frac{Q}{1+Q}) 
 \eeqn
 where $Q\sim (1+\Delta)^{3/2} e^{-x_f \Delta} $.  If $\Delta << 1$
  then $Q=O(1)$ and an efficient co-annihilation of 
XWIMPS can occur.   A detailed analysis was carried out in Ref.\cite{Feldman:2006wd}
using the packages of Ref.\cite{micro,Djouadi:2002ze} and 
 of Ref.\cite{DS,Paige:2003mg}. 
In the analysis  we have also examined
the effects of the $Z'$ pole on the relic density 
using the  techniques of  
integration over the $Z'$ pole in thermal averaging\cite{gs,Nath:1992ty}. The result  of the 
 analysis\cite{Feldman:2006wd}  is  exhibited in Fig.(3) using the constraints of the
three year WMAP data\cite{Spergel:2006hy} and using the parameter space of mSUGRA
accessible at the LHC\cite{lhc}.  Further, in
  the analysis of Ref.\cite{Feldman:2006wd} the  sensitivity to the top mass \cite{Nath:1995eq}
  was also investigated and the analysis found  to have significant  variations 
 with a one $\sigma$  variation in the top mass. 

 \noindent
Acknowledgements: 
The work was supported in part by
the U.S. National Science Foundation under the grant
NSF-PHY-0546568.

\bibliographystyle{ws-procs9x6}
\bibliography{ws-pro-sample}

\end{document}